\documentstyle[11pt,newpasp,twoside]{article}
\def\edcomment#1{\iffalse\marginpar{\raggedright\sl#1\/}\else\relax\fi}
\reversemarginpar

\begin{document}
\title{Extended Extragalactic Radio Emission}
 \author{Frazer N. Owen}
\affil{NRAO, P.O. Box O, Socorro, NM 87801 USA}
\author{Michael J. Ledlow}
\affil{University of New Mexico, Albuquerque NM USA}
\author{Jean A. Eilek}
\affil{New Mexico Tech, Socorro, NM 87801 USA}
\author{Namir E. Kassim}
\affil{NRL, Washington, DC USA}
\author{Neal Miller}
\affil{NRAO, P. O. Box O, Socorro, NM 87801 USA}
\author{K. S. Dwarakanath}
\affil{Raman Institute, Bangalore, INDIA}
\author{R. J. Ivison}
\affil{University College London, Gower Street, London WC1E 6BT UK}

\begin{abstract}
	Extended radio emission and its relation to parent galaxy
properties is briefly reviewed. Our current understanding of the relation 
between absolute radio and optical luminosity, radio morphology and 
linear size is discussed. The impact of radio jets on dense
cluster cores is discussed using M87 as an example. Finally, the relation of
AGN's to star-bursting galaxies at high redshift is considered.
\end{abstract}
\section{Introduction}

	Extended emission from radio galaxies is much too big a
subject to cover in a short talk or meeting proceedings. In this
contribution I will attempt to summarize some recent developments
on the properties of radio galaxies as a function of their environment
and their impact on that environment.

\section{Evolution of Extended Radio Emission}

	One important goal of the study of extragalactic radio sources
is to understand the life history of this phenomenon as a function of
time in the same way as we currently understand stellar evolution. It
seems likely that the evolutionary tracks of these sources depend on
several variables, including local environment and parent galaxy
properties. We can now begin to study these variables by looking at
fairly large samples of radio galaxies and studying the dependence
of the radio properties on the optical and X-ray data we can assemble
for them.  

	Given that most radio sources with luminosities with $L_{20cm} > 
 10^{23}$ W Hz$^{-1}$ ($H_0 = 75$ km s$^{-1}$ Mpc$^{-1}$) are believed to be 
formed by interactions between the jets and the external medium, one would think
that there would be differences in radio properties in and out of
cataloged rich clusters. However, studies of the statistical properties
of these two classes of sources has revealed little difference 
(Fanti 1984; Ledlow \& Owen 1996).  

       The lack of correlation of radio galaxy properties with rich
clusters is confusing. We can really only make this statement about
FR I's right now since the samples studied have been dominated by this
class. While it appears that clusters and their environments are not
important, another explanation is that the clusters are actually there
but we are missing them. Abell's sample of clusters, still the dominant
source list for these systems, is not based on a physical cutoff in
properties between ``clusters'' and the field but on the contrast between
these concentrations and the random galaxies along  the line-of-sight
considered. The probability of a galaxy being a radio source does not
depend on richness (Ledlow \& Owen 1995). We find numerous radio galaxies down into
Abell's richness class 0. Thus it might be reasonable to expect the cluster
population to continue down into lower richness systems.

        Recent X-ray and optical studies support this picture. Miller
el al (1999) and Hardcastle \& Worrall (1999) find evidence for both significant
optical clustering
for the Palomar DSS and diffuse X-ray atmospheres surrounding most of the nearby
FR I's studied with the short RASS exposures. Worrall (2000) also finds a
similar result using deeper, pointed ROSAT imaging. These results are
consistent with almost all FR I's being found in some sort of clustering with
an external hot atmosphere to confine the sources. However, lower richness
and size make the clusters harder to detect.

\begin{figure}
\plotfiddle{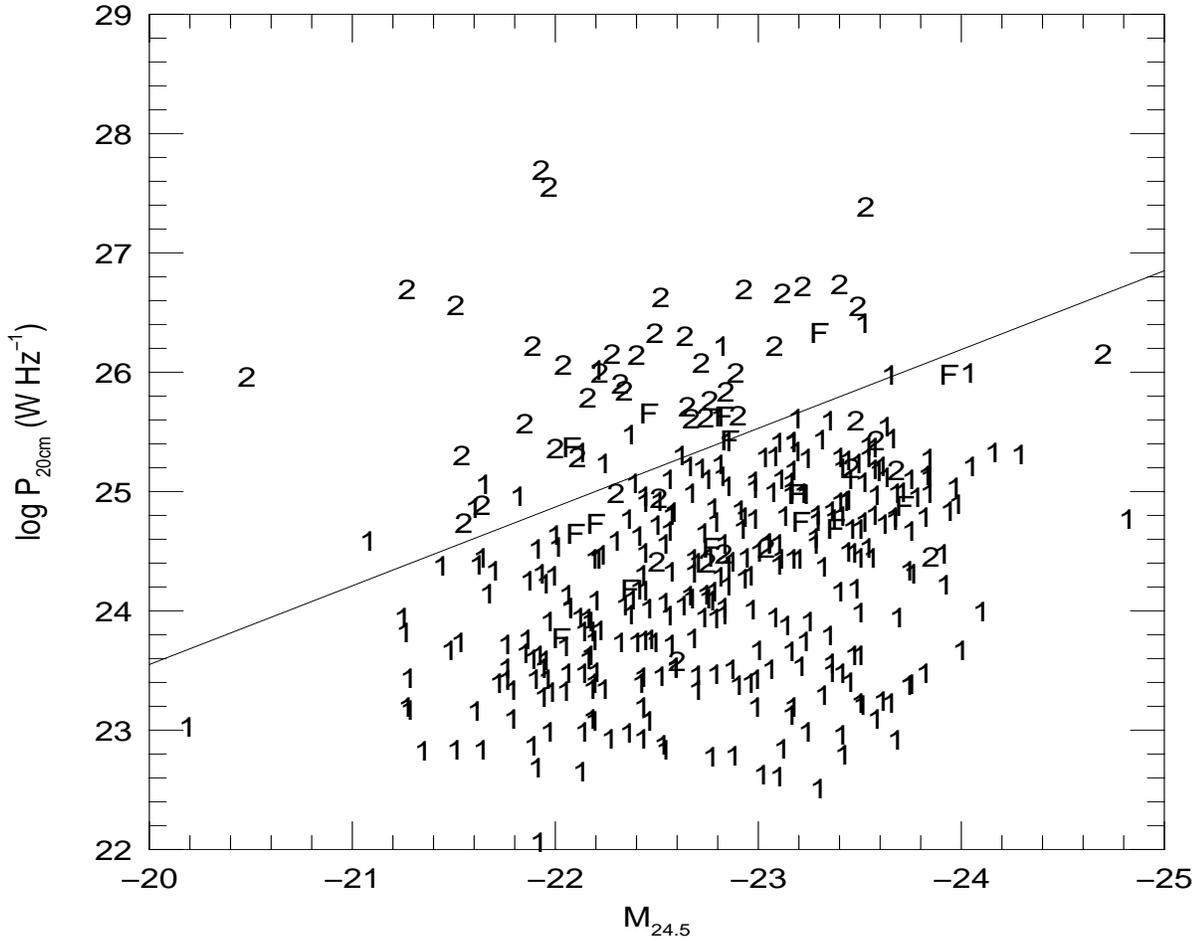}{3.2in}{0}{85}{70}{-270}{-100}
\caption{Correlation between Radio Luminosity, Optical Absolute Magnitude
and Fanaroff-Riley class. 1's are FR I's, 2's are FR II's and F's are
fat doubles. See Owen \& Laing (1989) and Ledlow et al (2000a,b) for details.}
\end{figure}

On the other hand, the shape of the
radio luminosity function seems to depend on galaxy absolute magnitude.
The FR I/II radio morphology also seems to depend on absolute magnitude in 
roughly the same way. That is, for each absolute magnitude, the radio 
morphology type changes at about
the same radio luminosity as the break occurs in the radio luminosity
function. This break is a strong function of optical absolute magnitude.
In figure 1 we show this relation for a recently completed
sample of radio galaxies taken from the Wall \& Peacock (1985) 2 Jy sample,
the Abell cluster sample and the Bologna radio galaxy sample. The
line shows the location of the break in properties as a function
of  absolute radio and optical luminosity. Please refer  to
Ledlow et al (2000a,b) for details.

        The correlation with optical galaxy properties may partly be
a result of the correlation of optical galaxy luminosity (and thus
presumably mass) with degree of clustering. More massive galaxies tend to
live in deeper, bigger potential wells and the properties of the
associated radio sources may depend primarily on this correlation. 
Investigators need to be careful that correlations of
properties of radio galaxies with redshift, including the degree of clustering,
are not a result of selecting objects from different parts of figure 1 as
a function of redshift.

	Our understanding of radio size as a function of radio power also 
gives us a hint about radio source evolution. (de Ruiter et al 1990, 
Ledlow et al 2000a,b).
There appears to be slow rise in the maximum linear sizes as a function of
radio power. However, for FR I's one worries that the picture may not
be complete. In figure 2, we show a new VLA image of B2 1108+27. Previous
observations of this source ($z=0.0331$) suggested a small FR I, $\sim 1$ 
arcmin in size (Fanti et al 1987). However, the NVSS image of this source 
suggested to us that there
might be a larger extension, although the ability to see this was at
or below the believable detection limit of the NVSS. The image with two 
hours integration with the VLA D-array reveals the 30 arcmin scale,
1 Mpc structure of this source.  

\begin{figure}
\plotfiddle{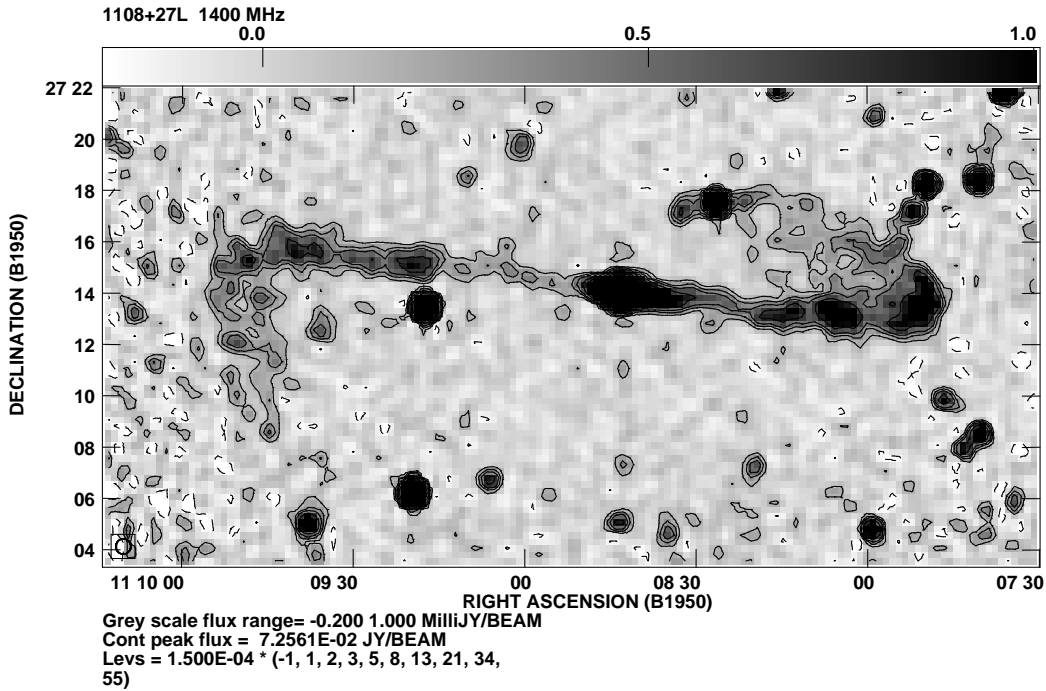}{3.5in}{-90}{55}{55}{-220}{300}
\caption{VLA image 1108+27 at 20cm with 2 hour integration}
\end{figure}

	The radio size versus optical and radio properties may be
more complicated and incomplete than the other relations. The large
sizes for many FR II's, and now FR I's, suggest very much larger diffuse
gaseous atmospheres than are usually detected by existing X-ray observations.
The present size diagrams, limited by surface brightness sensitivity, may
be part of the picture and may indicate that  characteristic scales for
the surrounding medium. But figure 2 may suggest
that this is not the complete picture and that we have much more to learn
about the evolution of individual radio galaxies and their lifetimes.

\section{Radio Galaxy Effects on the External Medium: M87}

	Many non-radio astronomers have often calculated the
absolute radio luminosity for a galaxy and then discounted its
importance given the usually much larger radiated luminosity
in other bands. However, most theoretical models of radio jets
suggest we are only seeing a small portion of the energy being
pumped out of the galaxy nucleus into the surrounding medium.
This energy input into the intracluster medium must be important
to the total energy budget of these systems. An example of this
can be seen in the recent VLA images of M87 at 90cm.

	In figure 3, we show the new 90cm VLA image (Owen, Eilek \&
Kassim 2000) of the full M87
system at 6 arcsec resolution and 50,000:1 dynamic range. Note
the large scale, $\sim 15$ arcmin. Most radio images to date have 
concentrated on the inner, much higher surface brightness part of
the source which contains the 20 arcsec long optical jet. The relationship
of the inner structure to the large-scale image shown here can be
studied on the web site (http://www.aoc.nrao.edu/$\sim$fowen/M87.html).
The new observations suggest that the flow seen in the inner part of
the source extends to much larger scales. The jet seems to be depositing
its energy on this scale and blowing up two bubbles in the process. The
radio emission appears to be distributed very non-uniformly in a complex
system of filaments some tens of kpc in length. The bubbles appear to have a
clear outer boundary which has the same maximum size over a broad range of
wavelengths from 4m to 20cm.

	The inner jet is modeled from a number of different points
of view as having a total kinetic energy flux of $ \sim 10^{44}$ ergs
s$^{-1}$ or more. However, the region occupied by the radio source has an X-ray 
luminosity of only $10^{43}$ ergs s$^{-1}$. This implies that more energy is
flowing into the region from the jet than out due to bremsstrahlung 
radiation. This is consistent with net heating of the volume and a
subsonic expansion of the bubbles. 

	This region has traditionally been modeled as a ``cooling flow''
using only the energy budget calculated from the X-ray emission. This
result suggests that the situation is currently much more complicated
for M87 at the center of the Virgo cluster. Of course, the radio event
may be transient on the time-scale of the cluster, so whether  the
cooling inflow or the jet (black hole) driven outflow dominates may
vary with time. The great variety of central radio sources in clusters
we see supports this qualitative picture. But
as with the radio jet sources, we need to study the
problem statistically to understand the evolutionary tracks of this
phenomenon.

\begin{figure}
\plotfiddle{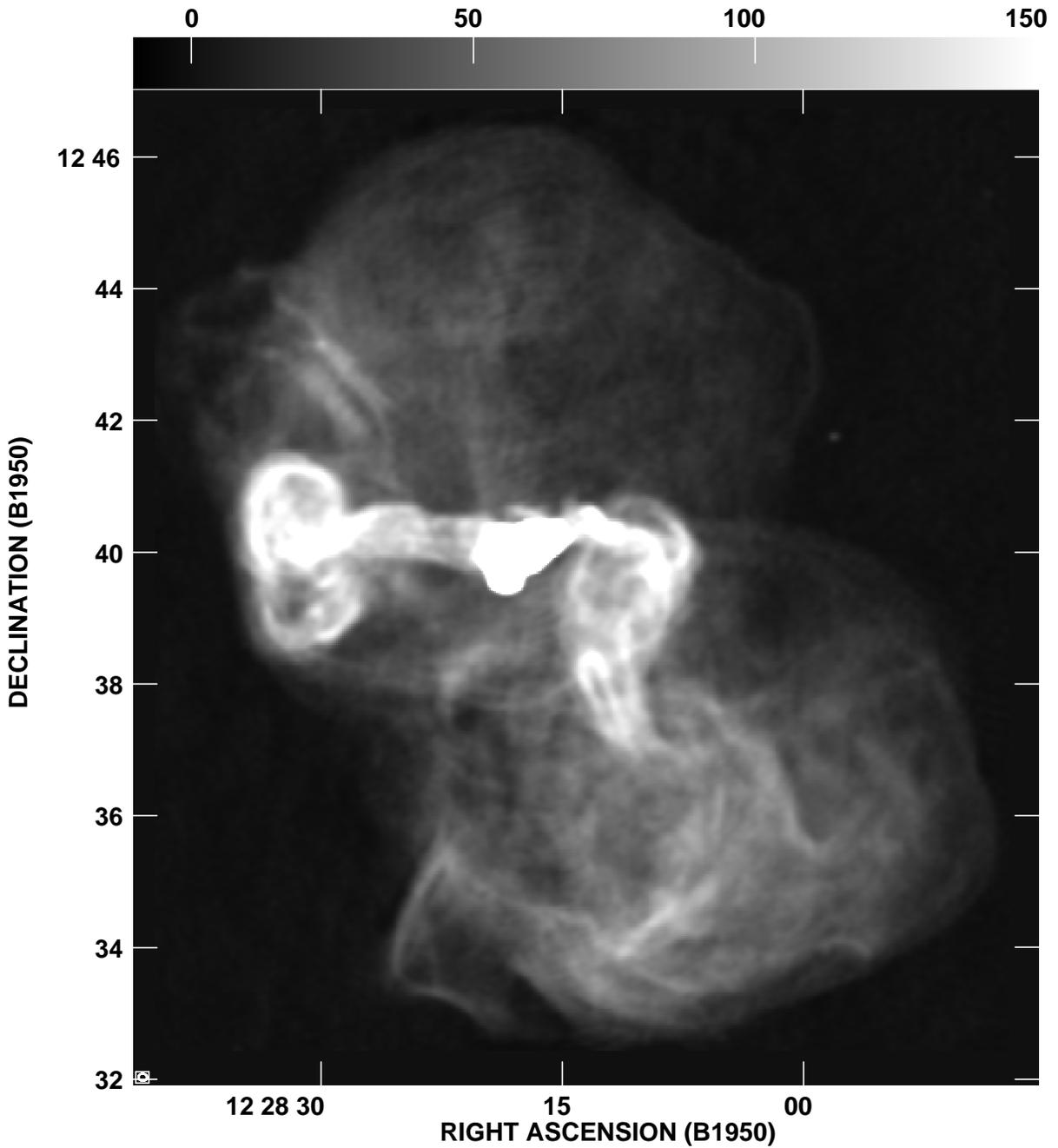}{6.5in}{0}{85}{85}{-250}{-100}
\caption{M87 image at 90cm. This image shows the structure of
the ``halo'' extending over 70 kpc. The saturated, triangular region near 
the center of the image contains the famous 2 kpc jet and the inner lobes.}
\end{figure}

\section{High-z Starformation and AGN's}

	Recently extragalactic, high-redshift dust emission has
been discovered using the SCUBA array on the JCMT. The positional
accuracy obtained for these weak sources with the relatively large
JCMT beam is limited. However, it turns out that many of the weak
submm sources are detected in deep 20cm surveys with the VLA as well.
The higher accuracy VLA positions can be used to identify the optical/IR
counterparts to these sources and when they are bright enough allow 
optical spectroscopy to determine their redshift and other properties.

	Although these sources have clear dust spectral signatures in
most cases, many of the identifications have turned out to have AGN
optical spectra. Also studies of nearby ultra-luminous FIR galaxies
have shown that often there is strong evidence for an AGN as well as
active star formation. This raises the important question of whether
the two processes (very high star-formation rates and active galactic
nuclei) are related in their origin, especially at high redshift.

	One of the first such objects discovered is JM$2399-0136$
(Ivison et al 1998). This
object has a fairly bright radio counterpart which allowed its confirmation.
Optically, it consists of two components separated by about 3 arcsec,
both with $z=2.8$. One of the objects has an AGN spectrum while the other
lacks the high excitation lines and thus is more consistent with 
star-formation. The B-array radio imaging shows a radio source with an
extent of 10 arcsec or more, consistent with an FR I morphology. Recently 
we imaged this field with the
VLA at 20 and 90cm in the A-array, producing images with 1.5 and 6 arcsec
resolution respectively. An earlier image at 1.5 arcsec resolution had been
obtained at 5 GHz. In the 20cm image we see two components which align with
the two optical objects. However, the one with the AGN spectrum appears to
have a much flatter spectrum typical of a normal, optically thin synchrotron
source. The lower excitation object appears to have a steeper spectrum and is
not detected at 5 GHz. The whole source, based on the 90cm results, appears
to have a spectrum steeper than one.

	At $z=2.8$, the spectrum of a synchrotron source in a weak magnetic
field should be steepened by Inverse Compton losses. This might well steepen
the spectrum of the low surface brightness parts of an FR I or the
synchrotron emission from the ISM of a galaxy. However, an AGN, perhaps
with a jet, could resupply the relativistic electron population rapidly
enough to maintain a normal spectrum. Thus this observation could suggest
that both processes are active in this object and that many starforming
sources at high redshift may have steep radio spectra. The extended, FR I-like
radio emission may mean that there is connection at high redshift between
radio jet sources and star-formation which we do not often see nearby.

\section{Conclusions}

	There are many exciting and important areas which the new low frequency
capabilities such as the GMRT can address and make major progress. These
are allowing us to ask questions about radio sources like those we
have posed for years about stellar evolution. Also, when combined with other
wavelength bands, we can begin see the connection and importance of
the phenonema we study to the rest of extragalactic astronomy and our very 
broad understanding of the universe.

	N.K. acknowledges basic research in radio astronomy at the Naval 
Research Laboratory is supported by the Office of Naval Research.

\end{document}